\documentclass[conference]{IEEEtran}
\IEEEoverridecommandlockouts
\usepackage[sort, compress]{cite}
\usepackage{amsmath,amssymb,amsfonts}
\usepackage{algorithmic}
\usepackage{graphicx}
\usepackage{textcomp}
\usepackage{xcolor}
\def\BibTeX{{\rm B\kern-.05em{\sc i\kern-.025em b}\kern-.08em
    T\kern-.1667em\lower.7ex\hbox{E}\kern-.125emX}}

\usepackage{float} 

\usepackage{hyperref}
\hypersetup{colorlinks=true}

\usepackage{listings}
\definecolor{codegreen}{rgb}{0,0.6,0}
\definecolor{codegray}{rgb}{0.5,0.5,0.5}
\definecolor{codepurple}{rgb}{0.58,0,0.82}
\definecolor{backcolour}{rgb}{0.95,0.95,0.92}
\usepackage[T1]{fontenc}
\usepackage{textcomp}

\begin{document}

\title{High-Performance Multi-Mode Ptychography Reconstruction on Distributed GPUs
\thanks{This work is supported by the Brookhaven National Laboratory's Laboratory Directed Research and Development project \#17-029. This work used resources of the Center for Functional Nanomaterials, of the Computational Science Initiative, and of the 3-ID Hard X-ray Nanoprobe Beamline of the National Synchrotron Light Source II, all belonging to a U.S.\ DOE Office of Science User Facility, at Brookhaven National Laboratory under Contract No.\ DE-SC0012704. \IEEEauthorrefmark{3}Equal contribution.}
}

\author{
	\IEEEauthorblockN{
		Zhihua Dong\IEEEauthorrefmark{1}\IEEEauthorrefmark{3}, 
   		Yao-Lung L.\ Fang\IEEEauthorrefmark{1}\IEEEauthorrefmark{2}\IEEEauthorrefmark{3}, 
		Xiaojing Huang\IEEEauthorrefmark{2}, 
		Hanfei Yan\IEEEauthorrefmark{2},\\
		Sungsoo Ha\IEEEauthorrefmark{1},
		Wei Xu\IEEEauthorrefmark{1},
		Yong S.\ Chu\IEEEauthorrefmark{2},
        Stuart I.\ Campbell\IEEEauthorrefmark{2},
		and Meifeng Lin\IEEEauthorrefmark{1}} 
	\IEEEauthorblockA{\IEEEauthorrefmark{1}Computational Science Initiative,\\
Brookhaven National Laboratory, Upton, NY 11973, USA}
	\IEEEauthorblockA{\IEEEauthorrefmark{2}National Synchrotron Light Source II,\\
Brookhaven National Laboratory, Upton, NY 11973, USA}
}


\maketitle

\begin{abstract}
Ptychography is an emerging imaging technique that is able to provide wavelength-limited spatial resolution from specimen with extended lateral dimensions. As a scanning microscopy method, a typical two-dimensional image requires a number of data frames. As a diffraction-based imaging technique, the real-space image has to be recovered through iterative reconstruction algorithms. Due to these two inherent aspects, a ptychographic reconstruction is generally a computation-intensive and time-consuming process, which limits the throughput of this method.
We report an accelerated version of the multi-mode difference map algorithm for ptychography reconstruction using multiple distributed GPUs. This approach leverages available scientific computing packages in Python, including mpi4py and PyCUDA, with the core computation functions implemented in CUDA C. 
We find that interestingly even with MPI collective communications, the weak scaling 
in the number of GPU nodes
can still remain nearly constant. 
Most importantly, for realistic diffraction measurements, we observe a speedup ranging from a factor of $\mathbf{10}$ to $\mathbf{10^3}$ depending on the data size, which  reduces the reconstruction time remarkably from hours  to typically about 1 minute and is thus critical for real-time data processing and visualization. 


\end{abstract}

\begin{IEEEkeywords}
X-ray ptychography, GPU, CUDA, MPI, Python
\end{IEEEkeywords}

\section{Introduction}

X-ray ptychography 
has a resurgence of interest in late 2000s due to the increase of both brilliance and coherence in modern synchrotron light sources \cite{PfeifferNatPho17}. Inheriting the sophisticated techniques and advantages from both coherent diffraction imaging (CDI) and scanning transmission X-ray microscopy (STXM),
it has become an essential tool in X-ray imaging at nanometer scale, and has the ability to produce high-resolution sample images while mitigating requirements for sample preparation, data analysis, optics, \emph{a priori} knowledge of probe, etc \cite{PfeifferNatPho17}. By scanning an X-ray probe over an extended specimen, one could measure the intensity of far-field transmission diffraction for each point, sometimes referred to as a ``view'', on a 2D scanning grid. However, phase information of the diffraction cannot be directly measured.  In order to reconstruct both the amplitude  and the phase for the light profile (probe) and specimen (object), 
overlapped scanning spots are necessary for providing sufficient information so that iterative algorithms can be used to retrieve the diffraction phase. 

There are several phase-retrieval algorithms in X-ray ptychography, including the ptychographical iterative engine (PIE) \cite{FaulknerPRL04,RodenburgAPL04} and its extension (ePIE) \cite{MaidenUlt09}, the difference map (DM) algorithm \cite{ThibaultSci08,ThibaultUlt09}, the non-linear optimization approach \cite{Guizar-SicairosOE08}, the maximum-likelihood optimization \cite{ThibaultNJP12}, etc.
In the present paper we focus on the DM algorithm, and implement it as part of the state-of-the-art multi-mode high-resolution tool suite at the Hard X-ray Nanoprobe (HXN) beamline at National Synchrotron Light Source II (NSLS-II) \cite{YanNF18}.

Without any parallelization, a typical ptychographic reconstruction of an image can take hours to process on a single CPU core of a standard workstation. Furthermore, in order to reduce the data acquisition time, most of the measurements are conducted in the ``on-the-fly'' scan mode \cite{PelzAPL14,DengOE15,HuangSR15} --- the sample is continuously moving relative to the probe. For accommodating the blurriness caused by continuous motion in the recorded diffraction data, multiple illumination modes have to be included in the reconstruction, which is introduced in \cite{ThibaultNat13}. This multi-mode approach increases both  memory footprint and computation time for the ptychographic reconstruction, making it crucial to have a high-performance ptychography reconstruction software.  

In this paper, we report the steps we took to port the HXN ptychography reconstruction software to distributed GPUs and the resulting performance improvements. The new GPU version reduces the computation time significantly down to only tens of seconds for reconstructing a single image with one or more probe/object modes, thereby fundamentally changing the workflow in the HXN beamline and providing real-time feedback to the facility users for arranging or adjusting the experimental setup.

\section{X-ray ptychography: difference map algorithm}
The DM algorithm is one of the most widely used and powerful phase-retrieval algorithms, which allows simultaneous reconstruction of the probe and the object, and our implementation follows closely  the original work \cite{ElserACSA03,ThibaultSci08,ThibaultNat13}. 
As far as we know this is the first reported GPU acceleration effort in the DM algorithm. Similar efforts utilizing GPUs include \cite{NashedOE14,MandulaJAC16,MarchesiniJAC16,NashedPCS17}.

For the fly scans,
the key insight is that the resulting blurriness can be equivalently seen as caused by an incoherent illumination beam.
As a result, the imperfection can be compensated by allowing the presence of multiple probe and object modes in the reconstruction. 
First, in the far-field limit the recorded transmission intensity is written as an incoherent sum of multiple modes, labeled by $k$ for probe and $l$ for object, respectively:
\begin{equation}
	I_j(\mathbf{q}) = \sum_{k,l} \left| \mathcal{F} \left[\psi_j^{(k,l)}(\mathbf{r})\right]\right|^2,
	\label{eq: intensity mode}
\end{equation}
where $\mathbf{r}$ is the position vector in the specimen plane and $\mathbf{q}$ the corresponding reciprocal vector, $\psi_j$ is the complex exit wave produced at the $j$-th scanning point and $\mathcal{F}$ stands for Fourier transform. In the following, it is assumed $j\in[1, N]$, $k\in [1, M_P]$, and $l\in[1, M_O]$. 
Next, we require $\psi$ to be expressed as a product of the $k$-th probe mode and $l$-th object mode,
\begin{equation}
\psi^{(k, l)}_j(\mathbf{r}) = P^{(k)}(\mathbf{r}-\mathbf{r}_j) O^{(l)}(\mathbf{r}), 
\label{eq: field product mode}
\end{equation}
where 
$\mathbf{r}_j$ is the scan position. 
The underlying assumption 
is that the probe variation along the propagation direction is negligible within the sample thickness.
The goal of ptychography is to reconstruct $P$ and $O$ iteratively subject to \eqref{eq: intensity mode} and \eqref{eq: field product mode}, 
which can be regarded as a constrained search in the hyperspace of complex pixels.

In the first step, one provides an initial guess of $P$ and $O$ so that the initial fields  can be constructed for each view $j$ based on \eqref{eq: field product mode}.
Next, the fields $\{\psi_j\}$ are iteratively updated according to
\begin{equation}
\begin{aligned}
 \psi_j^{(k,l)}(\mathbf{r}) &\leftarrow \psi_j^{(k,l)}(\mathbf{r}) + \beta \left\{ \mathcal{F}^{-1}\circ\mathcal{F}_c \left[ 2P^{(k)}(\mathbf{r}-\mathbf{r}_j)O^{(l)}(\mathbf{r})\right.\right.\\
 &\left.\left.\quad- \psi_j^{(k,l)}(\mathbf{r}) \right] - P^{(k)}(\mathbf{r}-\mathbf{r}_j)O^{(l)}(\mathbf{r}) \right\},
  \label{eq: update field}
 \end{aligned}
\end{equation}
 where $\beta\in[0, 1]$ is a free parameter empirically adjusted to speed up convergence (usually we set $\beta=0.8$), and $\mathcal{F}_c$ refers to ``constrained'' Fourier transform, in which the computed amplitude is replaced by the measured one while the computed phase is kept, 
 \begin{equation}
 \mathcal{F}_c \left[\psi_j^{(k,l)}(\mathbf{r})\right] \equiv \sqrt{\frac{I_j(\mathbf{q})}{\sum_{k,l}\left|\tilde{\psi}_j^{(k,l)}(\mathbf{q})\right|^2}}
 \tilde{\psi}_j^{(k,l)}(\mathbf{q})
 \end{equation}
 with $\tilde{\psi} = \mathcal{F} \left[\psi\right]$.
 The probe and object are also iteratively updated as follows:
 \begin{align}
 P^{(k)}(\mathbf{r}) &\leftarrow \frac{\sum_l\sum_j \left[ O^{(l)}(\mathbf{r}+\mathbf{r}_j) \right]^*\psi_j^{(k,l)}(\mathbf{r}+\mathbf{r}_j)} {\sum_l\sum_j |O^{(l)}(\mathbf{r}+\mathbf{r}_j)|^2},  \label{eq: update probe}\\
 O^{(l)}(\mathbf{r}) &\leftarrow \frac{\sum_k\sum_j \left[ P^{(k)}(\mathbf{r}-\mathbf{r}_j) \right]^*\psi_j^{(k,l)}(\mathbf{r})} {\sum_k\sum_j |P^{(k)}(\mathbf{r}-\mathbf{r}_j)|^2}. \label{eq: update object}
 \end{align}
 The convergence of all quantities can be monitored by calculating the relative 2-norm (L2-norm) distance between iterations, for example,
 \begin{equation}
 \varepsilon_P^{(k)} = \sqrt{\frac{\int d^2\mathbf{r}\left| P^{(k)}_\text{new}(\mathbf{r}) -  P^{(k)}_\text{old}(\mathbf{r}) \right|^2}{\int d^2\mathbf{r}\left|P^{(k)}_\text{new}(\mathbf{r})\right|^2}}.
 \end{equation}
 In our experience, the DM algorithm does not need many iterations to converge and to reach satisfying resolution, although the number of required iterations is sample-dependent. 

In practice, to determine the best choice  for the number of probe modes ($M_P$), we increase $M_P$ incrementally in a few trial runs and examine whether the contribution of each mode to the total intensity is larger than at least 1\% (if not, then stop).
So far we are unaware of any situation in which $M_O>1$ is useful, but we keep this flexibility in the code for future extensions.


\section{Porting to GPUs: strategy and implementation details}

We implemented the multi-mode DM algorithm on both CPU and GPU.
The original CPU code was written in Python for easy integration with other data acquisition, analysis, and visualization tools provided at NSLS-II. Therefore, in our port to NVIDIA GPUs we use PyCUDA (wrapper for CUDA API) \cite{PyCUDA}, scikit-cuda (for cuFFT and other CUDA libraries) \cite{scikit-cuda} and MPI for Python (mpi4py) \cite{mpi4py1,mpi4py2,mpi4py3} to accelerate the existing Python code. 
In addition, most of the computation are rewritten as CUDA C kernels, which are called through the PyCUDA binding for the best performance. An example of such binding is shown  below:
\begin{lstlisting}
import pycuda.driver as cuda
import pycuda.gpuarray as gpuarray
import pycuda.autoinit
import numpy as np

# assuming kernel "fx1" is implemented in sample.cu:
# extern "C" { // avoid C++ name mangling
#   __global__ void fx1(float* input) { 
#     /* code */ 
#   } 
#   /* other kernels */
# }

# load the CUDA binary (.cubin) generated by
# nvcc -cubin -o sample.cubin sample.cu
gpu_func_mod = cuda.module_from_file("sample.cubin")

# obtain a handle to the CUDA kernel
kernel_fx1 = gpu_func_mod.get_function("fx1")

# copy an array to GPU (a bit slower than htod)
a = gpuarray.to_gpu(np.arange(100, dtype=float))

# launch the kernel as if it were a Python function
# block and grid mean the same as in CUDA C
kernel_fx1(a, block=(100,1,1), grid=(1,1,1))
\end{lstlisting}

We first identified the hotspots in the code by coarse-grained performance profiling of the different function calls in the algorithm.  We found that in the serial implementation, the update of the views $\{\psi_j\}$ consumes the most CPU time and was our first target for GPU acceleration. Once this part has been ported to the GPUs, we redid the profiling and identified subsequently the error estimation for $\{\psi_j\}$, the update of the probe and object, and so on, as the next target for acceleration. 

Because GPU memory is scarce (ranging from a few to 32 GB depending on the device model), the full computation with realistic data size is usually difficult to be carried out using one GPU alone, especially for the multi-mode case as discussed earlier. It is therefore necessary to distribute the computation to multiple GPUs, either on the same compute node or across multiple nodes. 
Furthermore, in order to carry out most of the computation in the device and avoid large data transfer, 
we need to allocate extra GPU memory as buffer for FFT, product of $P$ and $O$, 
parallel reduction, etc.
As a result, assuming a single GPU can take a reconstruction of $N=40000$, probe size $200\times200$, and object size $10^3\times10^3$, the memory footprint is about $3$ times larger than the raw data. Another factor of roughly $3.7$ is needed if we have five probe modes ($M_P=5$). 
In total, over $10$ times of GPU memory compared to the raw data size is required for multi-mode ptychographic reconstruction\footnote{We note that cuFFT allocates additional memory that is not taken into account in our estimates.}, which 
can certainly be alleviated with multiple GPUs.

\begin{figure}[b]
	\centering
	\includegraphics[width=\linewidth]{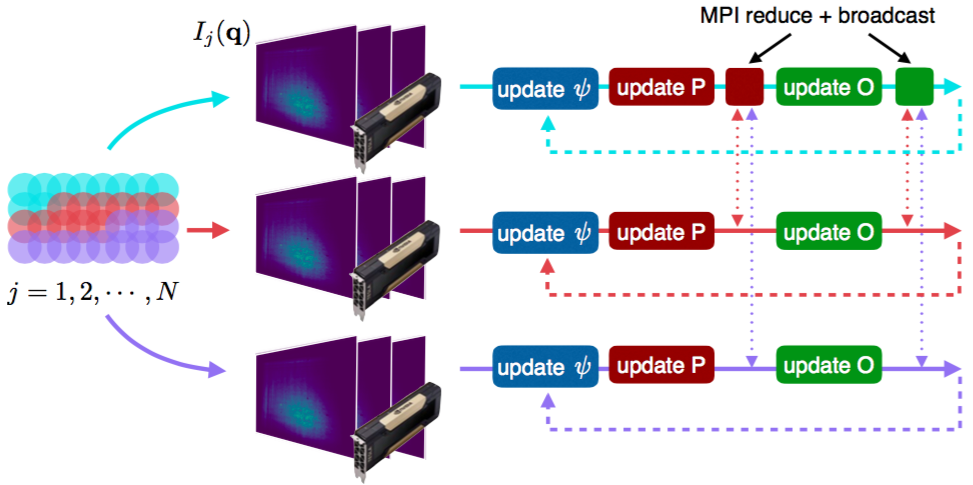}
	\caption{The workflow of our implementation. The diffraction data associated with $N$ scanning points are distributed (almost) evenly to available GPUs, along with initial guess of probe $P$ and object $O$, so that most of the computation is done in the device. At each iteration, 
	a few MPI collective communications (reduce and broadcast) are necessary to keep updated $P$ and $O$ visible across all GPUs.
	Other actions done in each iteration, such as error estimation, are not shown for clarity.
	}
	\label{fig: workflow}
\end{figure}


The workflow of our Python code is shown in Fig.~\ref{fig: workflow}. When the code starts, we divide (almost) equally the array indices of the dataset ($N$ intensity measurements)
by the number of available GPUs, possibly on different physical machines as long as reachable by MPI, let each GPU  grab its assigned portion from the disk, send  initial guesses for the probe and object to all GPUs, and start the iteration loop. 
Since each view $j$ is updated independently [cf.~\eqref{eq: update field}], this is a natural way to parallelize the computation,
and the consequence of such workload distribution is that the summation over $j$ in \eqref{eq: update probe} and \eqref{eq: update object} is partitioned into each GPU.
Therefore,
while each view $\psi_j$ is only updated in the GPU that it resides and no MPI collective communications are needed,
in each iteration we need to collect the (partially summed) probe and object  from each GPU, perform an MPI reduce followed by an MPI broadcast to keep the updated $P$ and $O$ visible to all GPUs.
We stress that because all views are needed to compute the probe and the object at each iteration, such synchronization is necessary, and we do not find this synchronization to be a significant burden to the entire calculation. Moreover, our approach avoids the complication of image stitching, such as those reported in \cite{NashedOE14}.  

One remark on the object reconstruction: Typically the probe dimension is much smaller than that of object. If we consider the functions of interest as matrices (since they live on a 2D plane labeled by pixel position $\mathbf{r}$), then effectively the object update according to \eqref{eq: update object} is an \emph{embedding} procedure:  the Hadamard product of two smaller matrices ($P$ and $\psi$) is embedded into a larger matrix ($O$). Such embedding is sketched as follows:
\begin{lstlisting}
# For each point j, the array indices (x_start and
# others) are calculated during initialization. Note 
# that the 1st dimension of prb and psi[j] is equal
# to x_end-x_start, and similarly for the 2nd dim
for j, (x_start, x_end, y_start, y_end) in \
    enumerate(point_info):
    obj_update[x_start:x_end, y_start:y_end] \
        += np.conjugate(prb) * psi[j]
    # perform other computations
\end{lstlisting}
Currently we perform this embedding in series (\textit{i.e.,} loop over scanning points $\mathbf{r}_j$) because such parallelization requires determining and grouping non-overlapping points in runtime, which is not trivial, and we are exploring possibilities of parallelizing it. As a workaround for mitigating this issue, we find that if the points on each GPU are batch-processed, then swapping the order of the $P$ and $O$ updates and then overlapping the $\psi$ update for batch $i+1$ with the $O$ update for batch $i$ can speed up by about 20\%. On the other hand, the probe counterpart \eqref{eq: update probe} does not have this problem as 
all pixels in $P$ are updated at once. 

Finally, we note that the initial guesses of $P$ and $O$ can be completely random or supplied by pre-existing results, and that initializing different probe modes with different degrees of  blurriness can help converge faster.

\section{Performance benchmark}
\begin{figure}[t]
	\centering
	\includegraphics[width=1.0\linewidth]{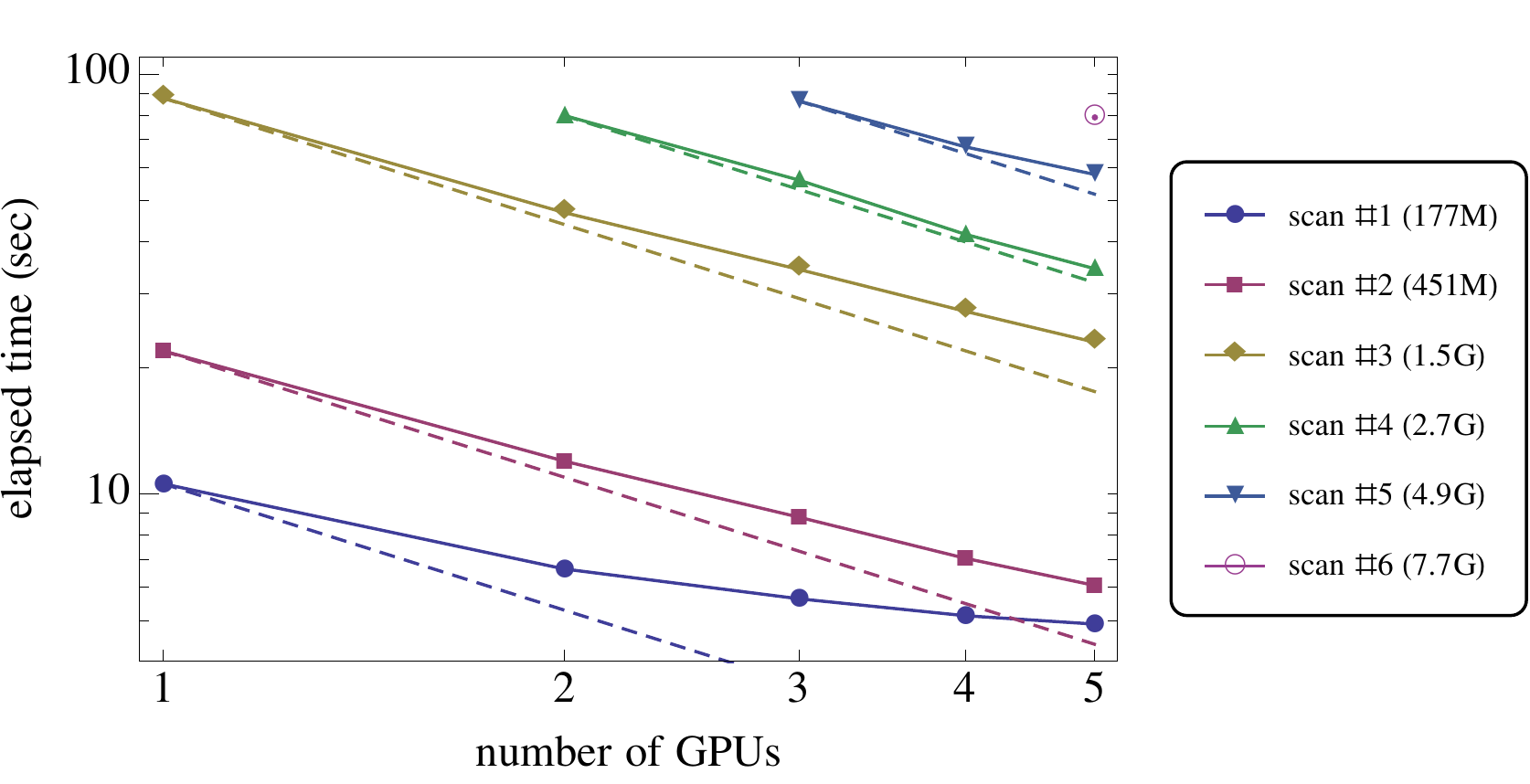}
	\caption{Strong scaling in the number of GPUs for completing 50 iterations for single-mode reconstruction, plotted in log-log scale. Real measurements of different $N$ and dimensions are used, and the raw data sizes are indicated. For some datasets (e.g.\ \#6) the data size is too large to fit in one GPU. Dashed lines are perfect scalings to guide the eyes. Tested on the HPC1 cluster in CSI, BNL. Each K20 compute node has Intel Xeon CPU E5-2670 @2.60GHz, 128GB physical memory, and one NVIDIA Tesla K20 GPU.
	}
	\label{fig: strong scaling}
\end{figure}

To benchmark the performance of our implementation, we first show in Fig.~\ref{fig: strong scaling} time-to-solution scaling as a function of the number of GPUs for realistic datasets with various sizes (different $N$ and small variation in probe dimension), a strong scaling in other words.
The speedup is most apparent for large datasets. In particular, for datasets of gigabytes order, the reconstruction can still be done on the order of a minute when using multiple GPUs. In fact, it can be seen that some datasets are too large to be fit in a single GPU (thus no data point on the plot), showing the necessity of using distributed GPUs. 

\begin{figure}[b]
	\centering
	\includegraphics[width=0.85\linewidth]{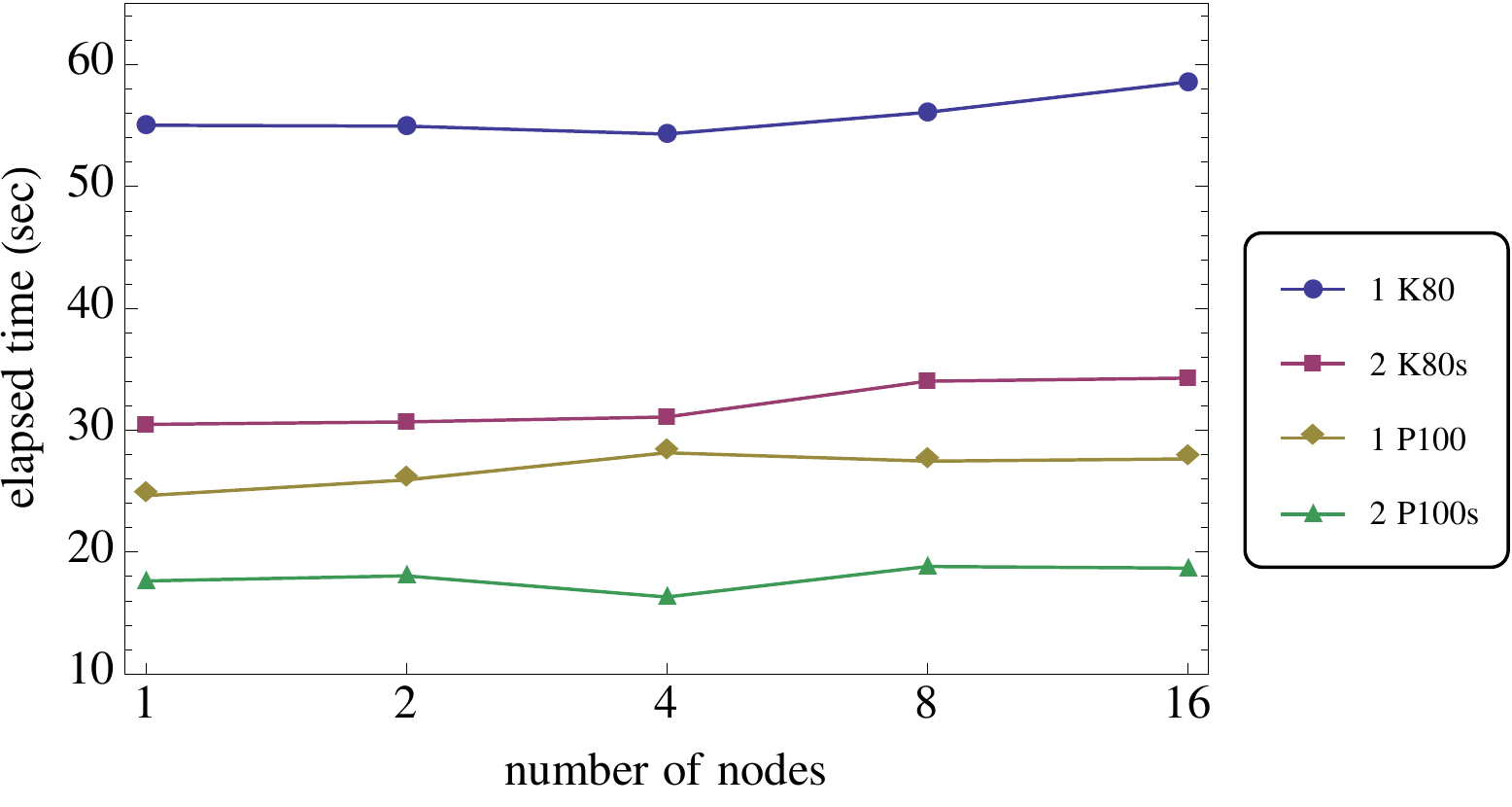}
	\caption{Weak scaling in the number of GPUs for completing 50 iterations for single-mode reconstruction, plotted in log-linear scale. Synthesized data of dimension $200\times200$ are used, and each node is assigned about 5000 points. Tested on the Institutional Cluster in BNL. Each compute node has Intel Xeon CPU E5-2695 v4 @2.10GHz, 256GB physical memory, Mellanox EDR InfiniBand connection, and either two NVIDIA Tesla K80 or two Pascal P100 GPUs.
	}
	\label{fig: weak scaling}
\end{figure}

Next, we simultaneously increase both the total problem size (the number of views $N$) and the number of GPUs while keeping the workload distributed to each GPU fixed, which is known as a weak scaling. 
In this case, we use instead synthesized data similar to the standard practice in the literature \cite{MaidenUlt09,HuangOE14,NashedOE14}:
we take two arbitrary images to be the object's amplitude and phase and generate the hypothetical diffraction pattern through a zone plate, with variable number of scanning points such that roughly $5000$ points are assigned to each compute node. The result is shown in Fig.~\ref{fig: weak scaling}, which shows a nearly constant scaling. We note that in this case due to the increasing communication overhead with more GPUs, the network infrastructure is critical to the scaling behavior. The measurement was done using the InfiniBand connection; with the TCP connection, we found that the scaling is less ideal but in general the elapsed time is still below 2 minutes for the highest $N$ (not shown).

\begin{figure}[t]
	\centering
	\includegraphics[width=\linewidth]{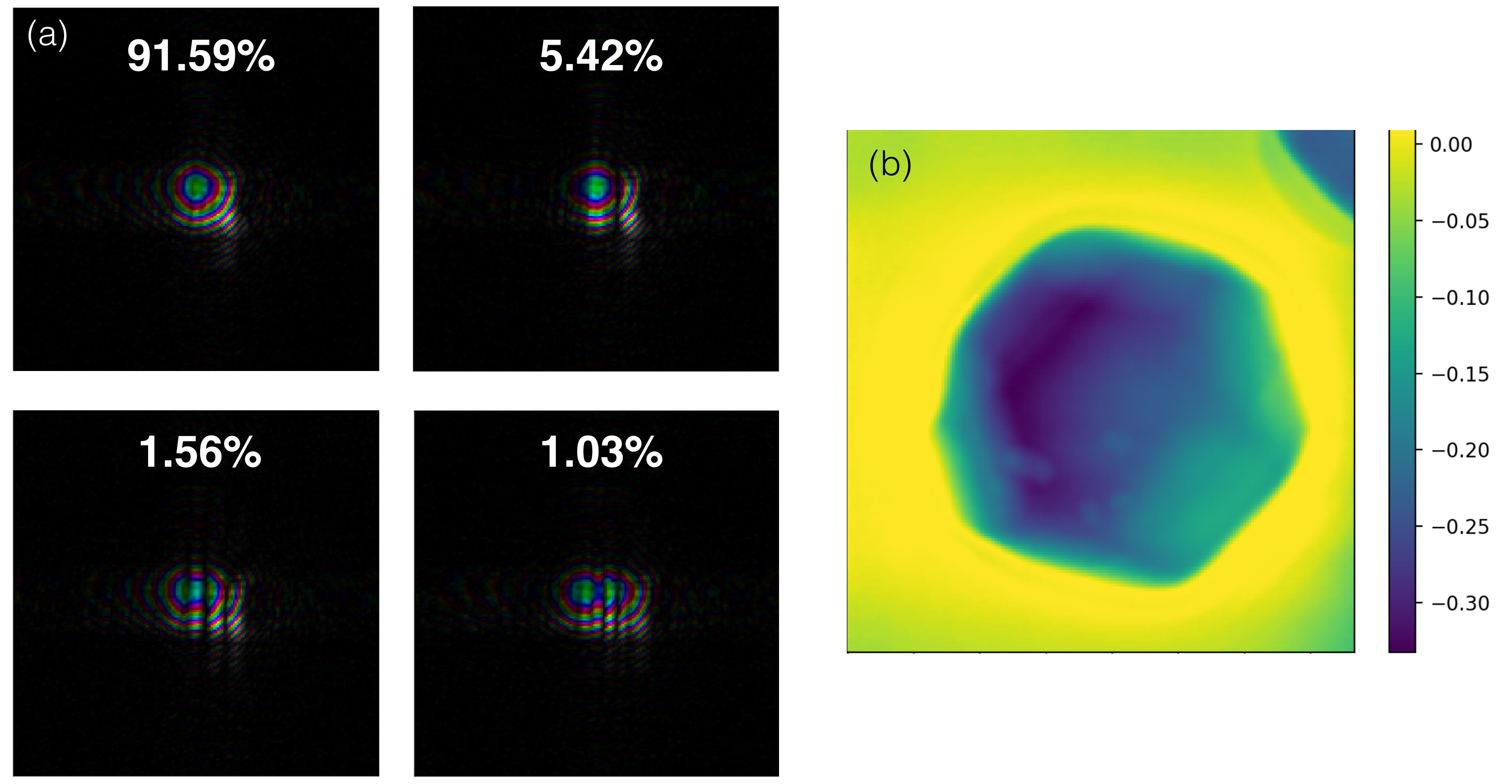}
	\caption{Multi-mode ptychographical reconstruction. (a) The first four dominant probe modes. (b) Reconstructed object phase, in which the field of view is $1\times 1 \mu$m. The sample is gold nano-crystals, prepared by annealing 20 nm thick gold film at 800$^{\circ}$C for 8 hours. Test machine: xf03id-srv5 in HXN, which has Intel Xeon CPU E5-2630 v4 @2.20GHz, 256GB physical memory, and four NVIDIA Tesla V100 GPUs\textsuperscript{\ref{HXNnote}}.
	}
	\label{fig: reconstruction}
\end{figure}

Finally, as a representative example, we present the results of a multi-mode reconstruction in Fig.~\ref{fig: reconstruction}. The raw measurement data is about 2.63GB ($N=10000$, image size $188\times188$, double precision), but since some scratch space is allocated by our code,
for this particular case more than 32 GB of memory is required if only one GPU were used, which is not possible even on the latest NVIDIA Tesla V100 product. Therefore, we use two and four V100 GPUs, and the corresponding timings for completing 50 iterations are 49.82s and 25.69s, respectively. Compared with the single-core CPU performance 
(8.8hr) on the same test machine,
the  speedup is about 1235x with 4 GPUs\footnote{The beamline machines have no job scheduler and can be accessed by all internal users, so a certain degree of performance degradation due to resource competition is expected\label{HXNnote}.}. 

\section{Conclusion and outlook}

In summary, we present a GPU implementation of single- and multi- mode DM algorithm for X-ray ptychography, which is already deployed at the HXN beamline of NSLS-II. The significant reduction of computation time from hours to tens of seconds is a game changer for the beamline scientists and users, allowing real-time feedback and shortening the analysis workflow. We emphasize that the GPU runtime is much shorter than the data acquisition time in diffraction measurements, therefore  the reconstruction can be done effectively in real-time as the measurement progresses.

We are continuing tuning the performance to further reduce the computation time and/or the memory footprint. For example, our preliminary tests indicate a slight improvement (about 10\% to 30\% speedup depending on the dataset size) on top of the achieved speedup by using page-locked host memory for data transfer, and we expect that this will benefit near-future experiments in which the object size is one order of magnitude larger in each dimension. Other possible  routes include using single-precision floating point arithmetic, adapting CUDA-aware MPI, overlapping kernel execution and data transfer, and parallelizing the object update as discussed earlier. 
We are also  porting other ptychographic algorithms to GPU as part of a high-performance ptychographic toolbox, 
all of which will soon be open-sourced  on \url{https://github.com/NSLS-II/}. 


\bibliographystyle{IEEEtran.bst}
\bibliography{Leo.bib}

\begin{thebibliography}{10}
\providecommand{\url}[1]{#1}
\csname url@samestyle\endcsname
\providecommand{\newblock}{\relax}
\providecommand{\bibinfo}[2]{#2}
\providecommand{\BIBentrySTDinterwordspacing}{\spaceskip=0pt\relax}
\providecommand{\BIBentryALTinterwordstretchfactor}{4}
\providecommand{\BIBentryALTinterwordspacing}{\spaceskip=\fontdimen2\font plus
\BIBentryALTinterwordstretchfactor\fontdimen3\font minus
  \fontdimen4\font\relax}
\providecommand{\BIBforeignlanguage}[2]{{%
\expandafter\ifx\csname l@#1\endcsname\relax
\typeout{** WARNING: IEEEtran.bst: No hyphenation pattern has been}%
\typeout{** loaded for the language `#1'. Using the pattern for}%
\typeout{** the default language instead.}%
\else
\language=\csname l@#1\endcsname
\fi
#2}}
\providecommand{\BIBdecl}{\relax}
\BIBdecl

\bibitem{PfeifferNatPho17}
F.~Pfeiffer, ``{X-ray ptychography},'' \emph{Nature Photonics}, vol.~12, no.~1,
  pp. 1--9, Dec. 2017.

\bibitem{FaulknerPRL04}
H.~M. Faulkner and J.~M. Rodenburg, ``{Movable Aperture Lensless Transmission
  Microscopy: A Novel Phase Retrieval Algorithm},'' \emph{Phys. Rev. Lett.},
  vol.~93, no.~2, p. 023903, Jul. 2004.

\bibitem{RodenburgAPL04}
J.~M. Rodenburg and H.~M.~L. Faulkner, ``{A phase retrieval algorithm for
  shifting illumination},'' \emph{Appl. Phys. Lett.}, vol.~85, no.~2, p. 4795,
  Nov. 2004.

\bibitem{MaidenUlt09}
A.~M. Maiden and J.~M. Rodenburg, ``{An improved ptychographical phase
  retrieval algorithm for diffractive imaging},'' \emph{Ultramicroscopy}, vol.
  109, no.~10, pp. 1256--1262, Aug. 2009.

\bibitem{ThibaultSci08}
P.~Thibault, M.~Dierolf, A.~Menzel, O.~Bunk, C.~David, and F.~Pfeiffer,
  ``{High-Resolution Scanning X-ray Diffraction Microscopy},'' \emph{Science},
  vol. 321, no.~5, pp. 379--, Jul. 2008.

\bibitem{ThibaultUlt09}
P.~Thibault, M.~Dierolf, O.~Bunk, A.~Menzel, and F.~Pfeiffer,
  ``\BIBforeignlanguage{English}{{Probe retrieval in ptychographic coherent
  diffractive imaging}},''
  \emph{\BIBforeignlanguage{English}{Ultramicroscopy}}, vol. 109, no.~4, pp.
  338--343, Mar. 2009.

\bibitem{Guizar-SicairosOE08}
M.~Guizar-Sicairos and J.~R. Fienup, ``{Phase retrieval with transverse
  translation diversity: a nonlinear optimization approach},'' \emph{Optics
  Express}, vol.~16, pp. 7264--, May 2008.

\bibitem{ThibaultNJP12}
P.~Thibault and M.~Guizar-Sicairos, ``{Maximum-likelihood refinement for
  coherent diffractive imaging},'' \emph{New J. Phys.}, vol.~14, no.~6, p.
  063004, Jun. 2012.

\bibitem{YanNF18}
H.~Yan, N.~Bouet, J.~Zhou, X.~Huang, E.~Nazaretski, W.~Xu, A.~P. Cocco,
  W.~K.~S. Chiu, K.~S. Brinkman, and Y.~S. Chu, ``{Multimodal hard x-ray
  imaging with resolution approaching 10 nm for studies in material science},''
  \emph{Nano Futures}, vol.~2, no.~1, p. 011001, Mar. 2018.

\bibitem{PelzAPL14}
P.~M. Pelz, M.~Guizar-Sicairos, P.~Thibault, I.~Johnson, M.~Holler, and
  A.~Menzel, ``{On-the-fly scans for X-ray ptychography},'' \emph{Appl. Phys.
  Lett.}, vol. 105, no.~2, p. 251101, Dec. 2014.

\bibitem{DengOE15}
J.~Deng, Y.~S.~G. Nashed, S.~Chen, N.~W. Phillips, T.~Peterka, R.~Ross,
  S.~Vogt, C.~Jacobsen, and D.~J. Vine, ``{Continuous motion scan ptychography:
  characterization for increased speed in coherent x-ray imaging},''
  \emph{Optics Express}, vol.~23, no.~5, p. 5438, 2015.

\bibitem{HuangSR15}
X.~Huang, K.~Lauer, J.~N. Clark, W.~Xu, E.~Nazaretski, R.~Harder, I.~K.
  Robinson, and Y.~S. Chu, ``{Fly-scan ptychography},'' \emph{Nature Scientific
  Reports}, vol.~5, p. 9074, Mar. 2015.

\bibitem{ThibaultNat13}
P.~Thibault and A.~Menzel, ``{Reconstructing state mixtures from diffraction
  measurements},'' \emph{Nature}, vol. 494, no. 7435, pp. 68--71, Jan. 2013.

\bibitem{ElserACSA03}
V.~Elser, ``{Solution of the crystallographic phase problem by iterated
  projections},'' \emph{Acta Crystallographica Section A}, vol.~59, pp.
  201--209, May 2003.

\bibitem{NashedOE14}
Y.~S.~G. Nashed, D.~J. Vine, T.~Peterka, J.~Deng, R.~Ross, and C.~Jacobsen,
  ``{Parallel ptychographic reconstruction},'' \emph{Optics Express}, vol.~22,
  no.~26, p. 32082, 2014.

\bibitem{MandulaJAC16}
O.~Mandula, M.~Elzo~Aizarna, J.~Eymery, M.~Burghammer, and V.~Favre-Nicolin,
  ``{PyNX.Ptycho: a computing library for X-ray coherent diffraction imaging of
  nanostructures},'' \emph{J Appl Crystallogr}, vol.~49, no.~5, pp. 1842--1848,
  Oct. 2016.

\bibitem{MarchesiniJAC16}
S.~Marchesini, H.~Krishnan, B.~J. Daurer, D.~A. Shapiro, T.~Perciano, J.~A.
  Sethian, and F.~R. N.~C. Maia, ``{SHARP: a distributed GPU-based
  ptychographic solver},'' \emph{J Appl Crystallogr}, vol.~49, no.~4, pp.
  1245--1252, Aug. 2016.

\bibitem{NashedPCS17}
Y.~S.~G. Nashed, T.~Peterka, J.~Deng, and C.~Jacobsen, ``{Distributed Automatic
  Differentiation for Ptychography},'' \emph{Procedia Computer Science}, vol.
  108, pp. 404--414, 2017.

\bibitem{PyCUDA}
A.~{Kl{\"o}ckner}, N.~{Pinto}, Y.~{Lee}, B.~{Catanzaro}, P.~{Ivanov}, and
  A.~{Fasih}, ``{PyCUDA and PyOpenCL: A Scripting-Based Approach to GPU
  Run-Time Code Generation},'' \emph{Parallel Computing}, vol.~38, no.~3, pp.
  157--174, 2012.

\bibitem{scikit-cuda}
L.~E. Givon, T.~Unterthiner, N.~B. Erichson, D.~W. Chiang, E.~Larson,
  L.~Pfister, S.~Dieleman, G.~R. Lee, S.~van~der Walt, B.~Menn, T.~M. Moldovan,
  F.~Bastien, X.~Shi, J.~Schl\"{u}ter, B.~Thomas, C.~Capdevila, A.~Rubinsteyn,
  M.~M. Forbes, J.~Frelinger, T.~Klein, B.~Merry, N.~Merill, L.~Pastewka, L.~Y.
  Liu, S.~Clarkson, M.~Rader, S.~Taylor, A.~Bergeron, N.~H. Ukani, F.~Wang, and
  Y.~Zhou, ``scikit-cuda 0.5.1: a {Python} interface to {GPU}-powered
  libraries,'' Dec. 2015, \url{http://dx.doi.org/10.5281/zenodo.40565}.

\bibitem{mpi4py1}
L.~Dalc\'{i}n, R.~Paz, and M.~Storti, ``{MPI for Python},'' \emph{Journal of
  Parallel and Distributed Computing}, vol.~65, no.~9, pp. 1108 -- 1115, 2005.

\bibitem{mpi4py2}
L.~Dalc\'{i}n, R.~Paz, M.~Storti, and J.~D'El\'{i}a, ``{MPI for Python:
  Performance improvements and MPI-2 extensions},'' \emph{Journal of Parallel
  and Distributed Computing}, vol.~68, no.~5, pp. 655 -- 662, 2008.

\bibitem{mpi4py3}
L.~D. Dalcin, R.~R. Paz, P.~A. Kler, and A.~Cosimo, ``{Parallel distributed
  computing using Python},'' \emph{Advances in Water Resources}, vol.~34,
  no.~9, pp. 1124 -- 1139, 2011, new Computational Methods and Software Tools.

\bibitem{HuangOE14}
X.~Huang, H.~Yan, R.~Harder, Y.~Hwu, I.~K. Robinson, and Y.~S. Chu,
  ``{Optimization of overlap uniformness for ptychography},'' \emph{Optics
  Express}, vol.~22, no.~10, p. 12634, 2014.

\end{thebibliography}

%
%
%

\end{document}